\documentclass{PoS}
\pdfoutput=1
\usepackage{rotating,booktabs,amsmath,amssymb, multicol,url}
\usepackage[numbers,sort&compress]{natbib}
\usepackage{natbibspacing}
\usepackage{ulem}
\usepackage{multicol}
\newcommand{\refcite}[1]{ref.~\cite{#1}}

\newcommand{\fig}[1]{figure~\ref{#1}}
\newcommand{\sect}[1]{section~\ref{#1}}

\newcommand{\be}{\ensuremath{\beta} }

\newcommand{\de}{\ensuremath{\de} }

\newcommand{\vev}[1]{\ensuremath{\left\langle #1 \right\rangle} }

\newcommand{\cN}{\ensuremath{\mathcal N} }
\newcommand{\cO}{\ensuremath{\mathcal O} }
\newcommand{\gc}{\ensuremath{g_c^2} }
\newcommand{\gGF}{\ensuremath{g_{\rm GF}^2} }

\newcommand{\gstar}{\ensuremath{g_{\star}^2} }
\newcommand{\gtc}{\ensuremath{\widetilde g_c^2} }
\newcommand{\gtGF}{\ensuremath{\widetilde g_{\rm GF}^2} }
\newcommand{\gtopt}{\ensuremath{\widetilde g_{\rm opt}^2} }
\newcommand{\topt}{\ensuremath{\tau_{\rm opt}} }
\newcommand{\MSbar}{\ensuremath{\overline{\mbox{MS}}} }

\newcommand{\bee}{\begin{equation}}
\newcommand{\ee}{\end{equation}}
\newcommand{\beea}{\begin{eqnarray}}
\newcommand{\eea}{\end{eqnarray}}

\title{Improved gradient flow for step scaling function and scale setting}

\ShortTitle{Improved gradient flow for step scaling function and scale setting}

\author{\speaker{Anna Hasenfratz}\\
        Department of Physics, University of Colorado, Boulder, CO 80309, USA\\
        E-mail: \email{anna@eotvos.colorado.edu}}

\abstract{The gradient flow renormalized coupling offers a simple and relatively inexpensive way to calculate the step scaling
function and the lattice scale, but both applications can be hindered by large lattice artifacts. Recently we
introduced an empirical non-perturbative improvement that can reduce, even remove $\mathcal{O}(a^2)$ lattice artifacts. The method is
easy to implement and can be applied to any lattice gauge theory of interest both in step scaling studies and
for scale setting. In this talk I will briefly review this improvement method and discuss its application for determining the discrete $\beta$ function of the   8
and 12 flavor SU(3) systems and for improved scale setting in 2+1+1 flavor QCD.}

\FullConference{The 32nd International Symposium on Lattice Field Theory,\\
		23-28 June, 2014\\
		Columbia University New York, NY}

\begin{document}

\section{Introduction}

The gradient flow renormalized coupling is a useful observable both   in scale setting and in step scaling function studies~\cite{Luscher:2010iy,Narayanan:2006rf}. It is easy to determine with high accuracy, however it    can suffer from large  discretization effects
~\cite{Borsanyi:2012zs,Sommer:2014mea,Fritzsch:2013je}.  While it is possible to systematically remove the lattice artifacts by improving  the lattice action, lattice operator, and the gradient flow force at the same time,  such improvement is not always practical~\cite{Ramos:2014kka}.  Tree level perturbative improvement developed in ~\refcite{Fodor:2014cpa} is easy to implement and appears to be effective at weak couplings. Recently we introduced the ``t-shift" improvement, an empirical approach to reduce, even remove the $\cO(a^2)$ corrections of the step scaling function.  ``t-shift" improvement is a non-perturbative method that can be used and appears to be effective  even at strong gauge couplings. In this talk I review this method and summarize results obtained for the discrete $\beta$ function in the $N_f=12$ and 8 flavor SU(3) systems~\cite{Cheng:2014jba,Hasenfratz:2014rna}. I also suggest a generalization of the ``t-shift" improvement  that can be used in scale setting and  illustrate it  using   Symanzik flow data on 2+1+1 flavor HISQ configurations~\cite{Bazavov:2013gca,Bazavov:2014sga}.

In \sect{sec:t-shift} I give a brief description of the t-shift improved gradient flow coupling,  following  closely Refs.~\cite{Cheng:2014jba,Hasenfratz:2014rna}. Section~\ref{sec:HISQ} illustrates the t-shift optimization of the $t_0$ gradient flow  scale, followed by the step scaling function study of the $N_f=12$ and 8 flavor systems in sections \ref{sec:Nf12} and \ref{sec:Nf8}.

\section{ \label{sec:t-shift}The t-shifted gradient flow running coupling}

The gradient flow renormalized coupling $\gGF$ at energy $\mu = \sqrt{8t}$   is defined  as
\begin{equation}
  \label{eq:def_g2}
  \gGF(\mu = 1 / \sqrt{8t}) = \frac{1}{\cN} \vev{t^2 E(t)},
\end{equation}
where  $t$ is the flow time, ${E}$ denotes the energy density, and the normalization constant $\cN$
is chosen such that $\gGF(\mu)$ agrees with the continuum \MSbar coupling at tree level.
The t-shift improved gradient flow coupling  introduced in \refcite{Cheng:2014jba} replaces 
     $\gGF(\mu; a)$   with the t-shifted coupling 
\begin{equation}
  \label{eq:t-shift}
  \gtGF(\mu; a) = \frac{1}{\cN} \vev{t^2 E(t + \tau_0 a^2)}\, ,
\end{equation}
where $\tau_0 \ll t / a^2$ is a small shift in the flow time.
In the continuum limit $\tau_0 a^2 \to 0$ and $\gtGF(\mu) = \gGF(\mu)$.
At finite lattice spacing $\gtGF(\mu)$ differs from $\gGF(\mu)$ by an $\cO(\tau_0 a^2)$ term. Since $\gGF(\mu)$ itself has $\cO(a^2)$ cut-off corrections, an appropriate choice of $\tau_0$ can remove those, making the t-shifted coupling $\cO(a^2)$ improved
\begin{equation}
  \gtopt(\mu; a) = \gGF(\mu; a = 0) + \cO(a^4 [\log a]^n,a^4).
\end{equation}

The t-shift improvement contains only a single parameter, $\tau_0$. To achieve full $\cO(a^2)$ improvement this $\tau_0$  must depend on other parameters, like $\gtGF(\mu)$ and  the bare gauge coupling.
Fortunately in practice it is sufficient to choose $\tau_0$ to be a constant or only weakly $\gtGF(\mu)$ dependent to remove most $\cO(a^2)$ lattice artifacts. The t-shift does not have to be fully optimized,  even if 
$\tau_0$ does not remove all $\cO(a^2)$ corrections, the t-shifted coupling has the correct continuum limit. 

The t-shift improved $t_\alpha$   lattice scale can be defined as 
$t_\alpha^2\langle E(t_\alpha+\tau_0)\rangle = \alpha$ .
 The traditional $t_0$  scale correspond to $\alpha=0.3$ but any other (reasonable) value can be used. Requiring that the relative scales $t_\alpha/t_0$ are independent of the lattice spacing, i.e. identical on  different configuration ensembles, defines the  optimal $\tau_0$ value as will be illustrated in \sect{sec:HISQ}.
 
 In  step scaling function studies the optimal $\tau_0$ removes the $\cO(a^2)$ terms of the discrete \be function \begin{equation}
  \label{eq:beta_lat}
  \be_{\rm lat}(\gc; s; a) = \frac{\gtc(L; a) - \gtc(sL; a)}{\log(s^2)}.
\end{equation}
The subscript $c$ in $\gc$ relates  the energy scale and the lattice size, $\sqrt{8t} = 1/\mu = c L$ ~\cite{Fodor:2012td}, its value is usually $c = 0.2 - 0.4$. 

Since the gradient flow is evaluated through numerical integration, the replacement $\gGF \to \gtGF$ can be done by a simple shift of $t$ without incurring any additional computational cost both in scale setting and step scaling studies. 


\section{\label{sec:HISQ}The t-shift improved lattice scales}

I illustrate the t-shift improved scale setting with 2+1+1 flavor HISQ ensembles~\cite{Bazavov:2010ru,Bazavov:2012xda} using  Symanzik flow data generated for  \refcite{Bazavov:2013gca}.
 I consider the ensembles with physical strange mass and three different light to strange quark mass ratios: $m_s/m_l=5$ (m1), 10 (m2) and 27 (m3). The ensembles, according to their lattice spacings, fall into four groups,  $a\approx0.15$fm, 0.12fm, 0.09fm and 0.06fm  (ensembles A, B, C and D). Further details can be found in \refcite{Bazavov:2013gca}.

\begin{figure}[btp]
  \centering
  \includegraphics[width=0.49\linewidth]{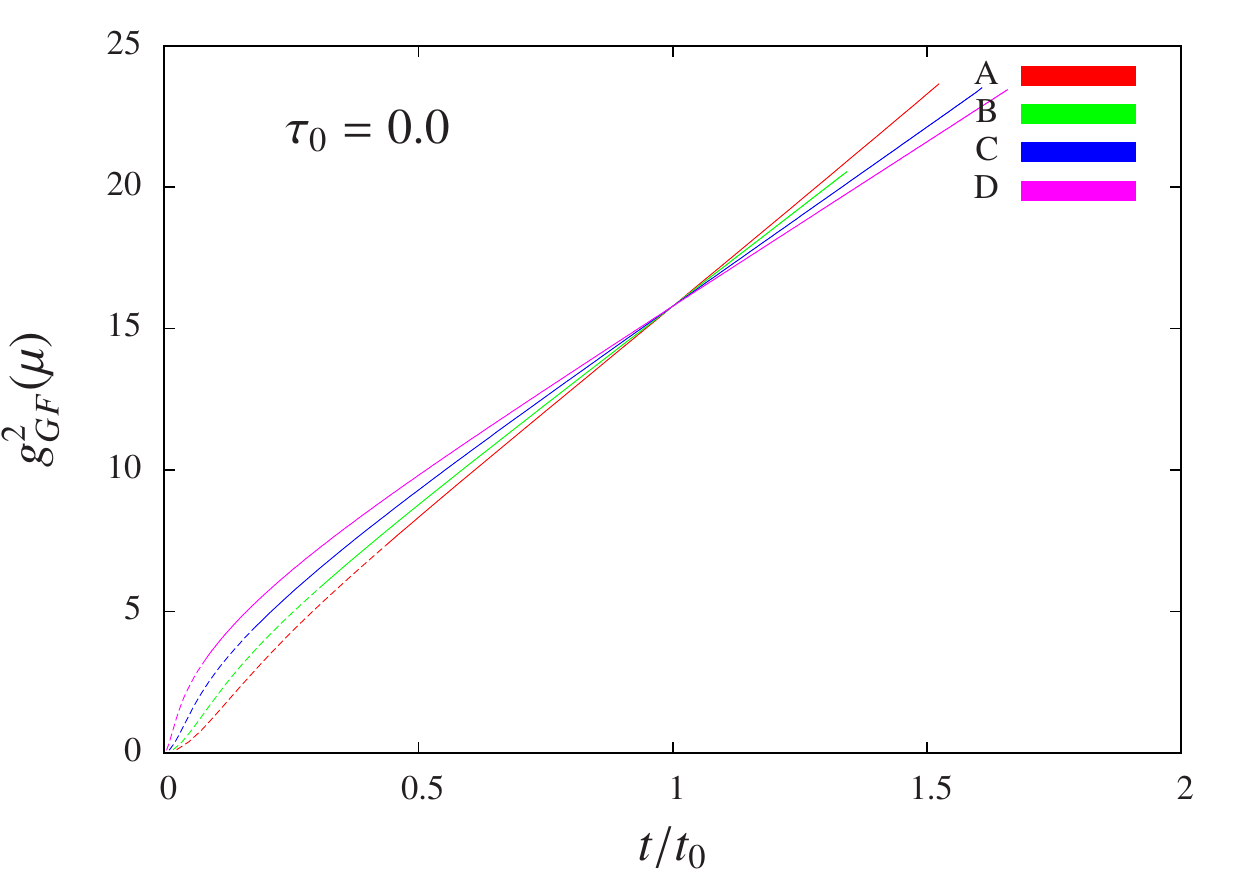}\hfill
   \includegraphics[width=0.49\linewidth]{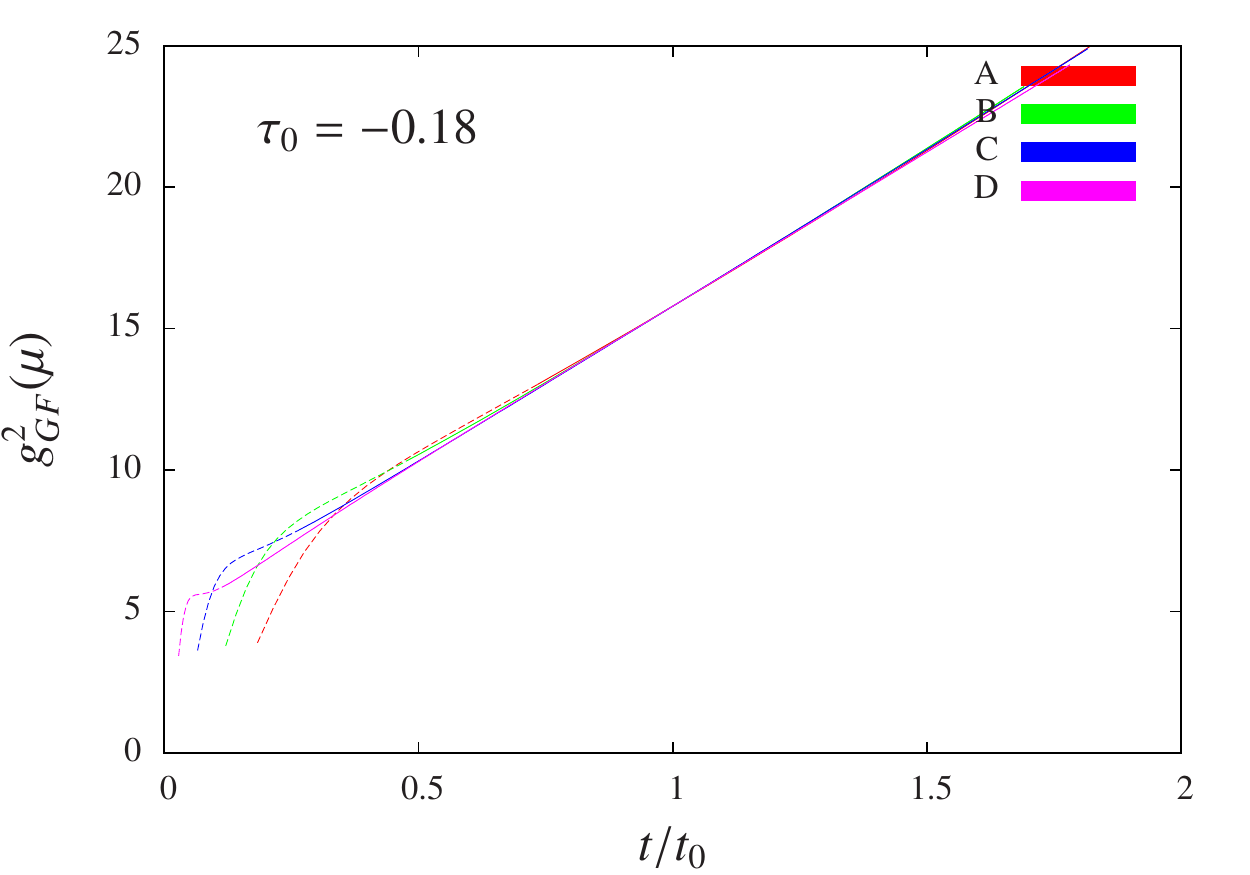}
  \caption{\label{fig:g2_vs_mu_00} Left panel: The gradient flow running coupling $\gGF(\mu)$  without t-shift improvement ($\tau_0=0$) of the $m_s/m_l=27$ (m3) data set as the function of  $t/t_0$.
    Errors are smaller than the width of the lines. The dashed lines signal where $\sqrt{8t}/a<2$ and lattice artifacts  could be large. Right panel: Same as the left panel but with near-optimal $\tau_0=-0.18$ t-shift improvement.  }
\end{figure}

The lattice artifacts  of $t_0/a^2$ for the $m_s/m_l=27$ data set are  illustrated in  the left panel of  \fig{fig:g2_vs_mu_00} where  I  compare the  gradient flow running coupling $\gGF(\mu;a)$ as the function of the flow time $t$ normalized by $t_0$   for the  four different lattice spacings ensembles.  Since the different curves are rescaled with their corresponding $t_0/a^2$ lattice scale,  they are forced to cross at $t/t_0=1$.

The condition $t^2\langle E\rangle =0.3$ to set the scale is arbitrary, any (reasonable) value of $\alpha$  
in the relation $t_\alpha^2\langle E(t_\alpha+\tau_0)\rangle = \alpha$ 
should predict consistent lattice scales. If the gradient flow scale  had no lattice spacing corrections, the curves corresponding to  different ensembles in the left panel of   \fig{fig:g2_vs_mu_00} would   overlap in a wide  range,  limited only by  finite volume effects at large  $t$ and by    gradient flow integration  effects at  small $t$. This is obviously not the case.

\begin{figure}[btp]
  \centering
  \includegraphics[width=0.49\linewidth]{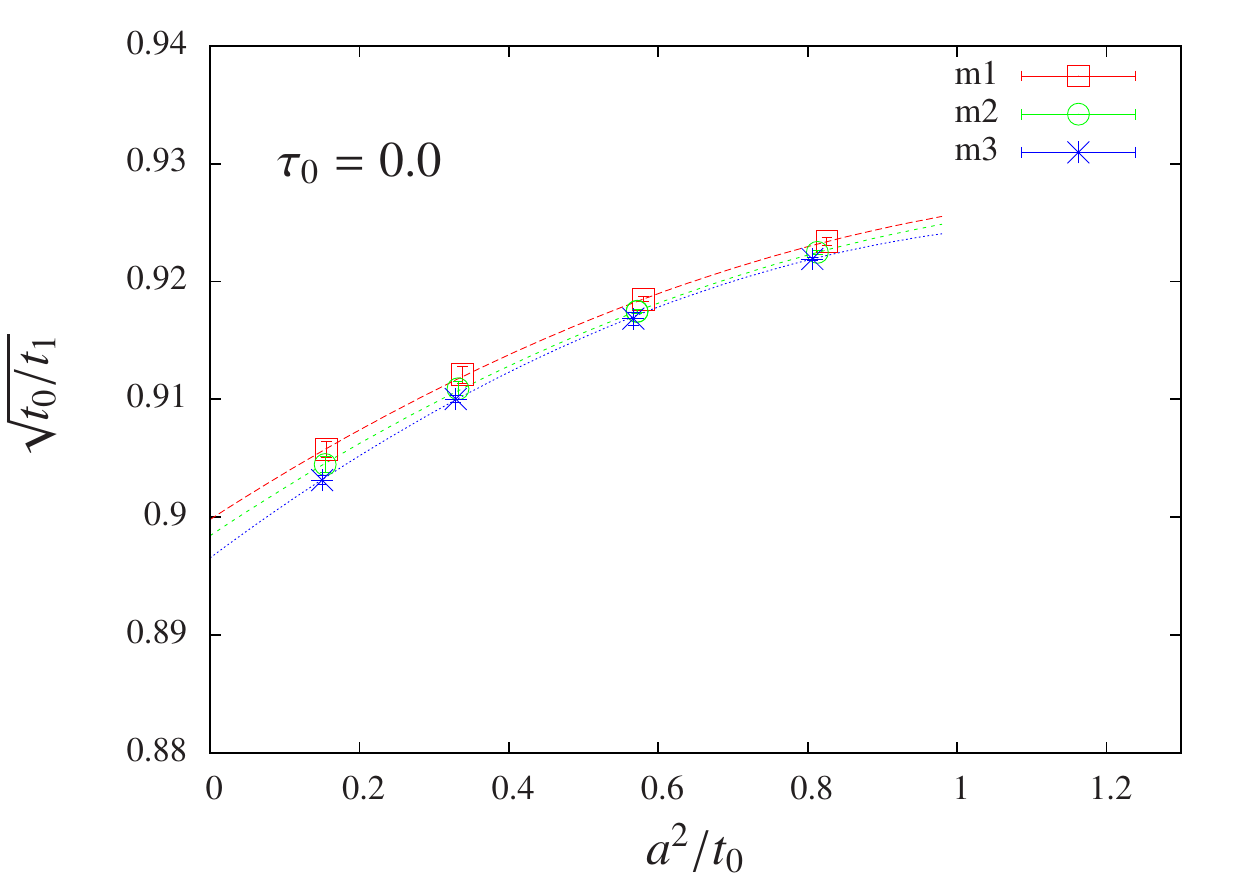}
   \includegraphics[width=0.49\linewidth]{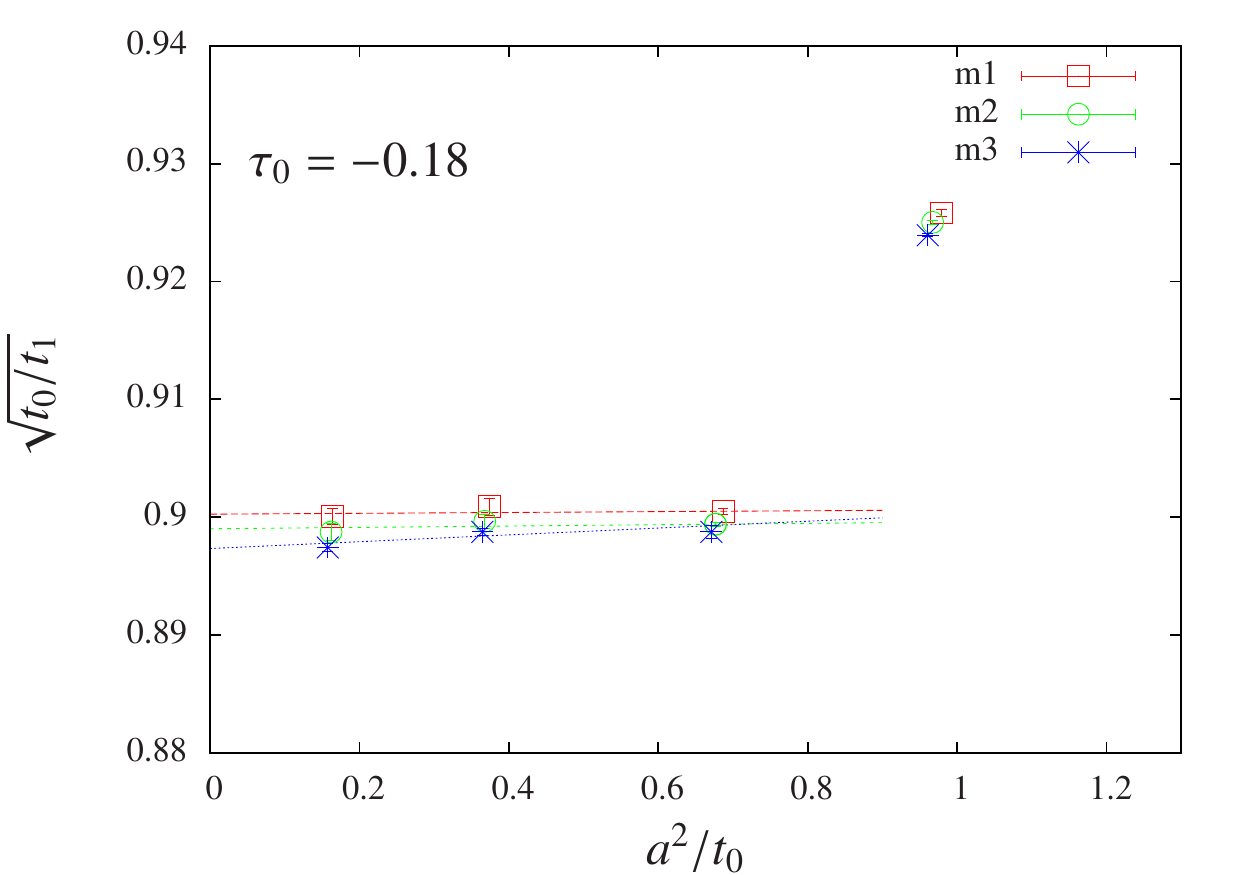}
  \caption{\label{fig:t0t1} $\sqrt{t_0/t_1}$  as the function of $a^2/t_0$. The different colors/symbols correspond to different $m_l/m_s$ ratios. Left: $\tau_0=0$, no t-shift improvement. The dashed lines are quadratic extrapolations. Right: $\tau_0=-0.18$ improved gradient flow. The dashed lines are linear extrapolations excluding the coarsest (A) configuration set.  }
\end{figure}

To quantify the lattice artifacts  I  compare two  different scales,   $t_0$  at $\alpha=0.3$  and $t_1$  at
 $\alpha=0.35$.
The left panel of \fig{fig:t0t1} shows the  ratio $\sqrt{t_0/t_1}$ as the function of the lattice spacing $a^2$  for all three $m_s/m_l$ data sets. This ratio would be constant if there were no lattice artifacts, but as the figure shows  $\sqrt{t_0/t_1}$ is not only not constant, it does not even show a  linear dependence on $a^2$.

 Introducing a small $\tau_0$ shift changes the predicted scales slightly and improves the scaling considerably. The right panel of  \fig{fig:t0t1}  shows the ratio $\sqrt{t_0/t_1}$ with $\tau_0=-0.18$. This value was obtained by trial-end-error without attempting to identify a precise optimal value: any $\tau_0$ is correct in the sense that it should predict the same continuum limit. 
 
 Since  I use the same $\tau_0$ value at every bare gauge coupling, I do not expect full $\cO(a^2)$ improvement. Nevertheless the  three finest  data sets predict a nearly constant value for $\sqrt{t_0/t_1}$ suggesting only minimal lattice artifacts. At the same time   the coarsest ``A" data set at $a\approx 0.15$fm shows significantly cut-off corrections and cannot be included in any reasonable continuum limit extrapolation.  
  It appears that 
 the coarse $a\approx 0.15$fm configurations  are not in the $\cO(a^2)$ scaling region of the quantity $t_0$. Without t-shift improvement the rather large lattice artifacts  mask this problem.

The right panel of  \fig{fig:g2_vs_mu_00}  is the analogue of the left panel  showing the 
  gradient flow running coupling now with near-optimal $\tau_0=-0.18$  t-shift for the m3 ensembles.  $\gtGF(\mu;a)$ shows little variation with the lattice spacing after the initial gradient flow integration artifacts die out, a significant improvement over the $\tau_0=0$ case in the left panel. 
  
\begin{figure}[btp]
  \centering
  \includegraphics[width=0.48\linewidth]{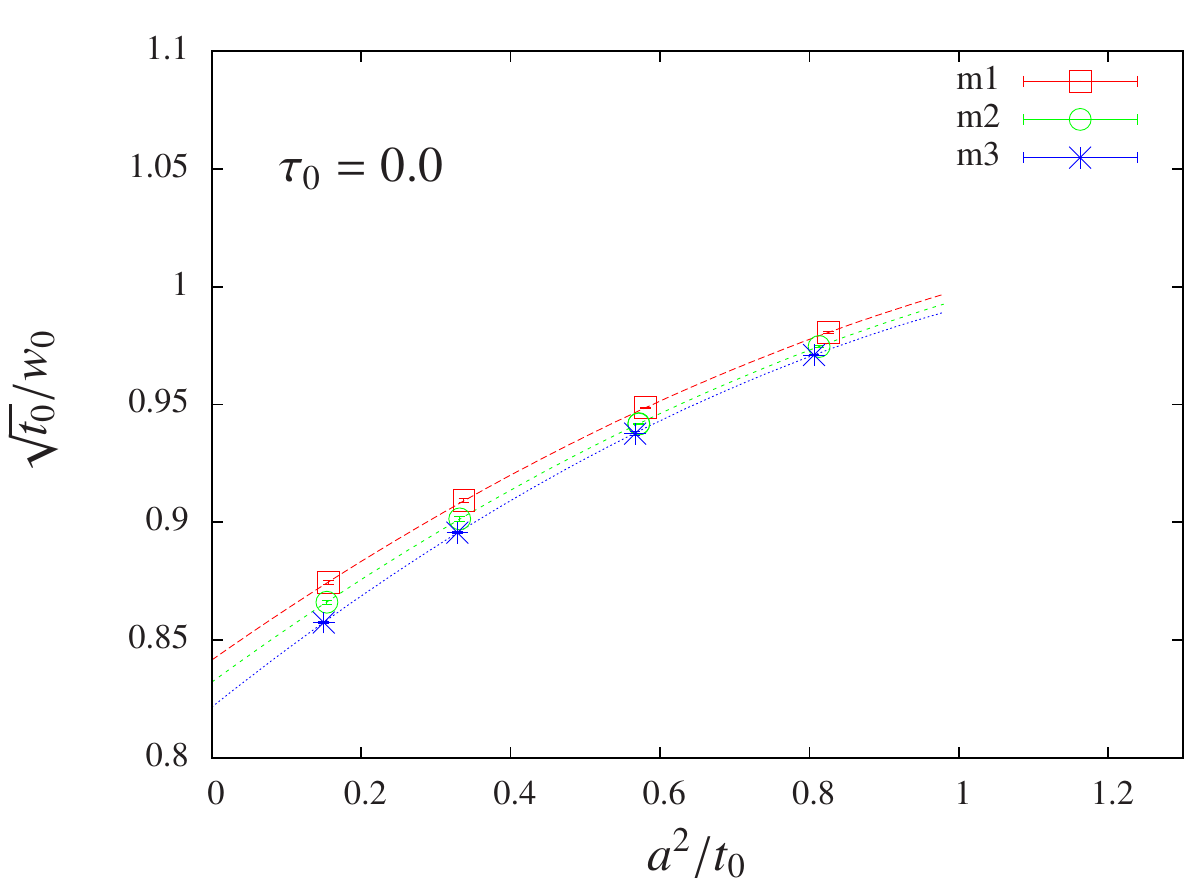}
   \includegraphics[width=0.5\linewidth]{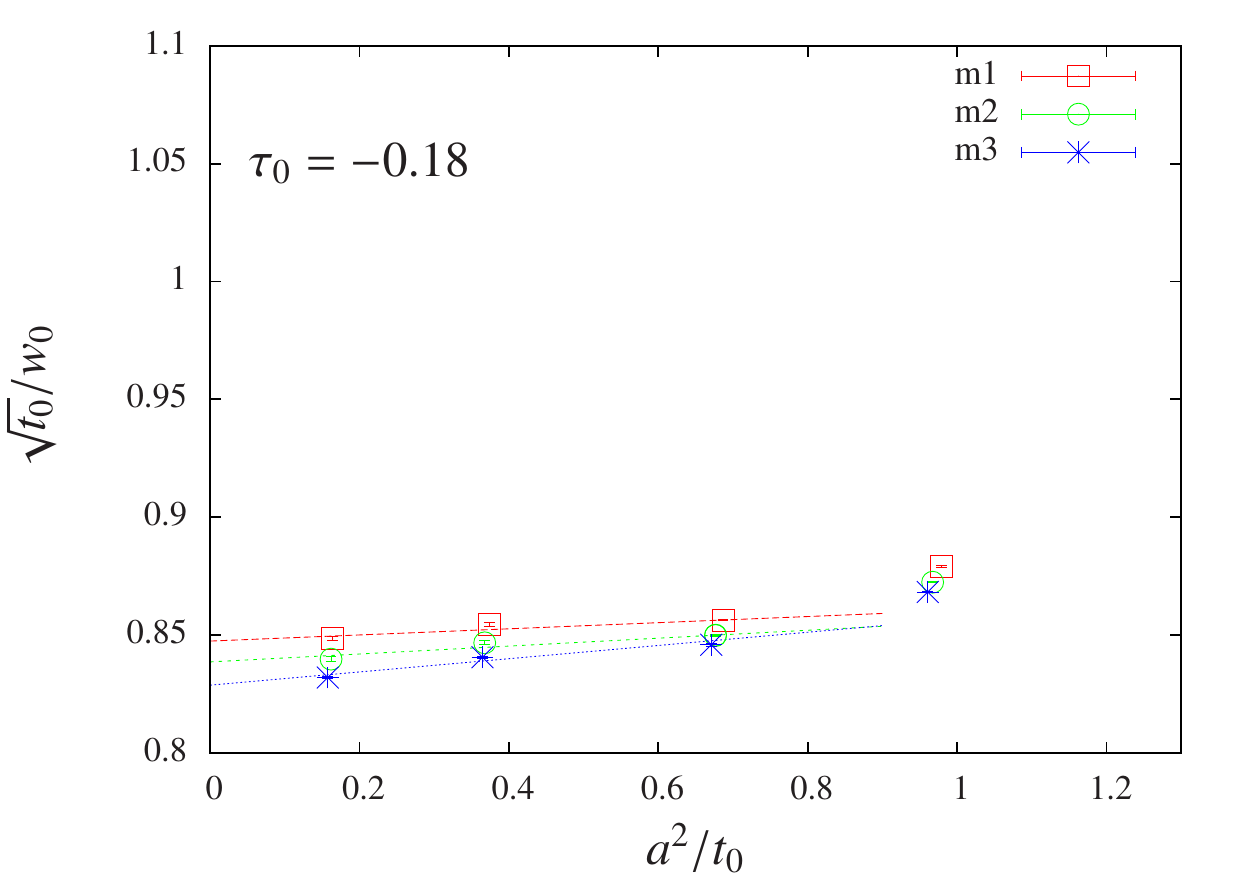}
  \caption{\label{fig:t0w0} Same as {\protect \fig{fig:t0t1}} but for the ratio $\sqrt{t_0}/w_0$. }
\end{figure}

 It is worth considering other lattice scales   to check lattice artifacts with and without the t-shift improvement. As an example  in \fig{fig:t0w0} I show  $\sqrt{t_0}/w_0$ with $\tau_0=0.0$ and -0.18.  Lattice corrections are significant without t-shift, while the  t-shifted ratio is almost constant and can be extrapolated linearly if only the three finer data sets are considered. As before, the coarsest  ``A" configuration sets do not appear to be in the $\cO(a^2)$ scaling regime. Comparing $t_0$ to the more conventional $r_1$ Sommer scale leads to very similar conclusions. 

In a recent study of 4 light and 8 heavy flavors we used the  gradient flow coupling in a similar way. We found that a small t-shift removed most observable lattice artifacts in the infrared, showing the emergence of walking behavior as the mass of the 8 heavy flavors decreased~\cite{Brower:2014ita}

\section{\label{sec:Nf12} $N_f=12$ flavors}

The12-flavor 3-color system has been investigated by several groups with somewhat controversial conclusions. While recent finite size scaling analysis suggests that the infrared behavior of this system is consistent with conformality and mass anomalous dimension $\gamma_m=0.235(5)$~\cite{Cheng:2013xha,Lombardo:2014pda}, a step scaling function study remains the most reliable method to identify its infrared fixed point. 

In \refcite{Cheng:2014jba} we used nHYP smeared staggered fermions  and considered six different volumes up to $36^4$. We  performed the simulations in the $m = 0$ chiral limit and
the range of volumes allowed us to carry out step scaling analyses with scale changes $s = 4 / 3$, $3 / 2$ and 2.
We chose the gradient flow coupling \gc with  $c=0.2$
to minimize the statistical errors, though we  verified that other $c$ values (0.25 - 0.35) gave consistent results.

\begin{figure}[btp]
  \centering
  \includegraphics[width=0.52\textwidth]{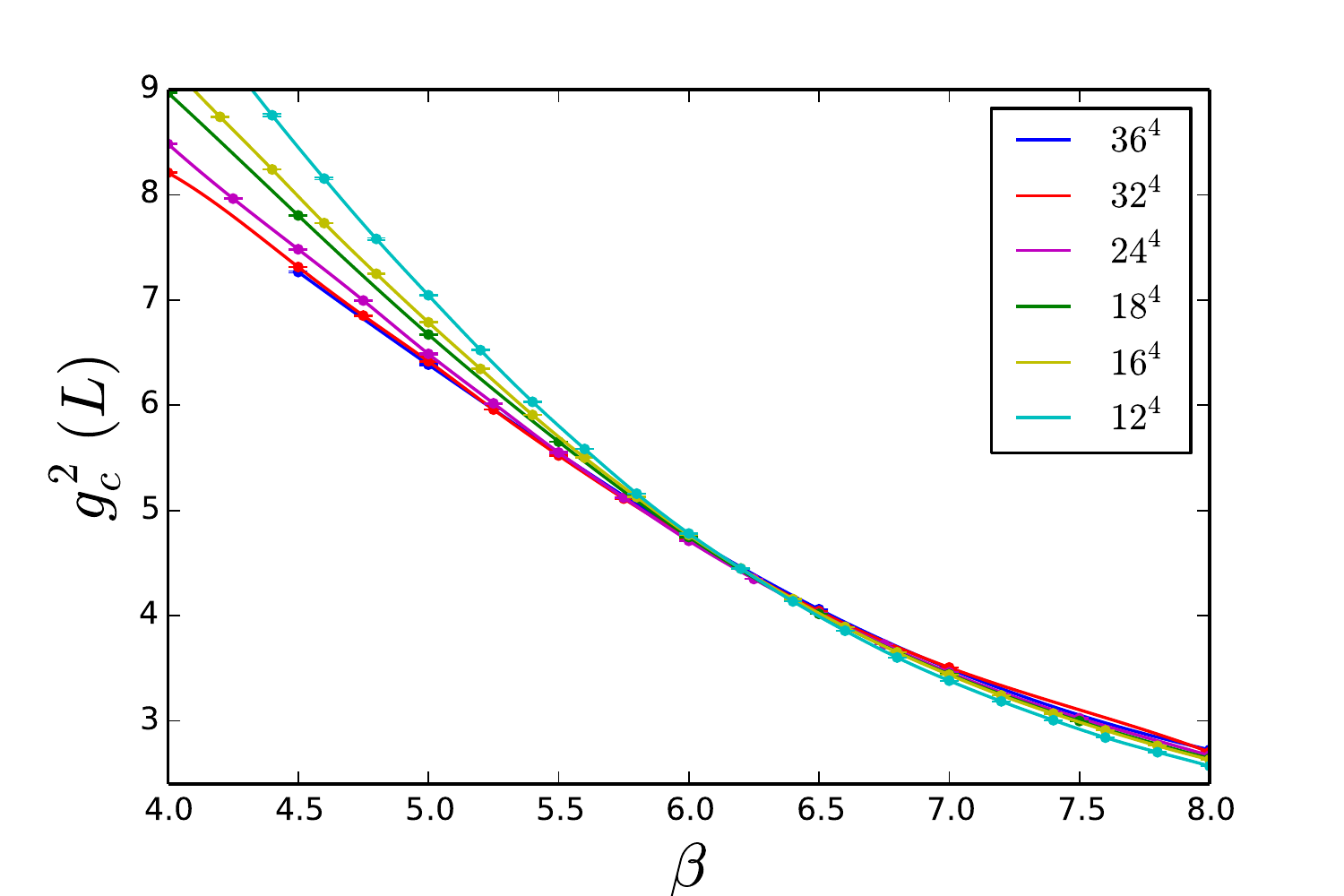}\hfill
  \includegraphics[width=0.45\textwidth]{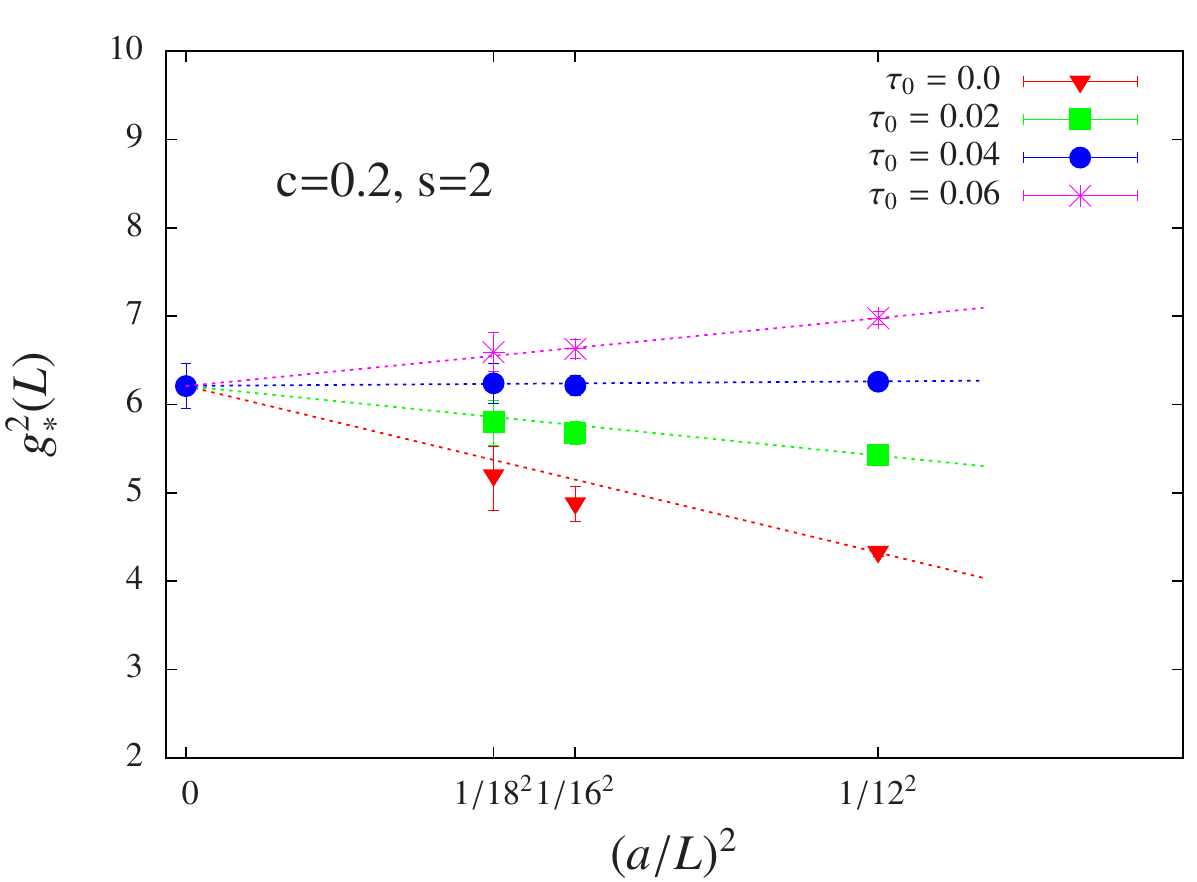}
  \caption{\label{fig:Nf12}Left panel:The $N_f = 12$ running coupling $\gc(L)$ versus the bare coupling $\be_F$ on several volumes, for $c = 0.2$.  
  Right panel:Continuum extrapolations of the 12-flavor finite volume IRFP $\gstar(L)$, with several different $t$-shift coefficients $\tau_0$ for fixed scale change $s = 2$.  
  }
\end{figure}

The left panel of ~\fig{fig:Nf12} shows the running coupling $\gc(L)$ as the function of the bare gauge coupling $\be_F$ for different volumes.
The interpolating curves are polynomial fits that 
 cross in the range $6.0 \leq \be_F \leq 6.5$.
The crossing from lattices with linear size $L$ and $sL$ defines the finite-volume IRFP coupling $\gstar(L; s)$.
If the IRFP exists in the continuum limit then the extrapolation
$ \lim_{(a / L)^2 \to 0} \gstar(L; s) \equiv \gstar$
has to be finite and independent of the scale change $s$.
The right panel of ~\fig{fig:Nf12} illustrates the continuum extrapolation of $\gstar(L)$ with scale change $s = 2$ for various choices of the $t$-shift parameter $\tau_0$.
Most  lattice cut-off effects are removed with  $t$-shift, $\topt \approx 0.04$.

The near-optimal $\topt \approx 0.04$
turns out to be  also near-optimal for scale changes $s = 3 / 2$ and $4 / 3$,
 making the extrapolation to the continuum very stable, predicting that the IR fixed point is  at renormalized coupling $\gstar = 6.18(20)$ in the $c = 0.2$ scheme.

\section{\label{sec:Nf8} $N_f=8$ fundamental flavors}

The $N_f=8$ flavor SU(3) system has received significant attention as a chirally broken system near the conformal window and   candidate model  for walking behavior. 2-loop perturbation theory predicts that $N_f=8$ is below the conformal window though at 3- and 4-loop the \MSbar $\beta$ function develops an infrared fixed point at strong coupling. In \refcite{Hasenfratz:2014rna} we studied the step scaling function of this model using two different  lattice actions, one with once the other with twice  nHYP smeared staggered fermions. 
The continuum extrapolated step scaling function must be independent of the lattice action so comparing different actions at identical renormalized couplings can give insight of systematical errors.

Our numerical simulations were carried out in the $m=0$ chiral limit. We did not observe spontaneous chiral symmetry breaking even though we probed fairly large renormalized couplings, reaching $\gtc(L) = 18.0(1)$ on our largest $30^4$ volume  with two nHYP smearing steps.

\begin{figure}[btp]
  \includegraphics[width=0.45\textwidth]{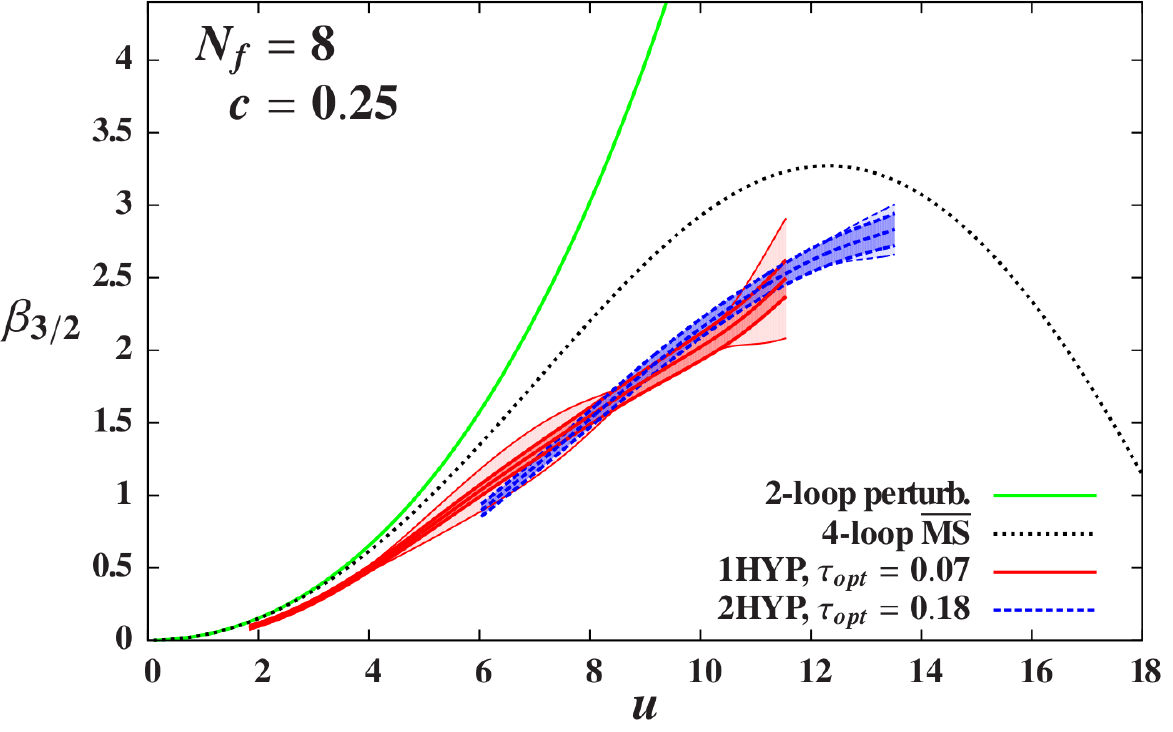}\hfill
  \includegraphics[width=0.45\textwidth]{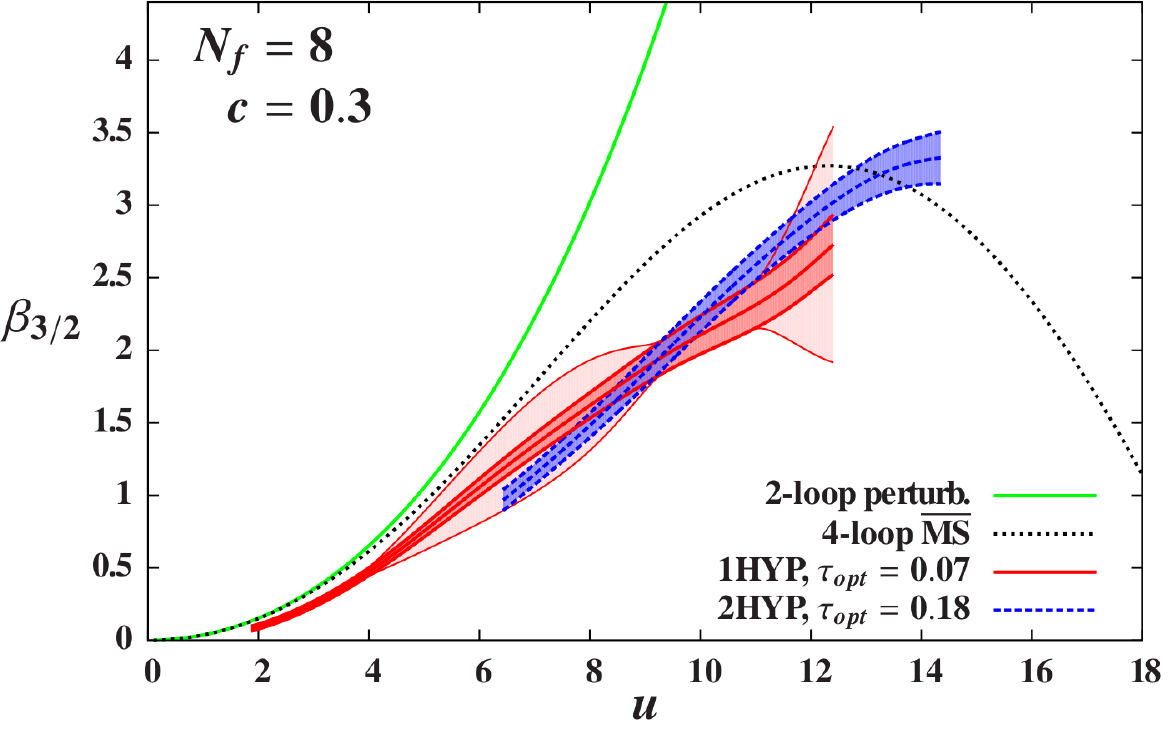}
  \caption{\label{fig:beta}Continuum-extrapolated discrete \be function for scale change $s = 3 / 2$ with $c = 0.25$ (left) and 0.3 (right).  In each plot we include once- and twice-smeared results using the optimal $\topt = 0.07$ and 0.18, respectively, as well as two-loop perturbation theory (solid line) and the four-loop perturbative prediction in the \MSbar scheme (dotted line).  The darker error bands indicate our statistical uncertainties, while the lighter error bands show the total uncertainties, with statistical and systematic errors added in quadrature.}
\end{figure}

Figure~\ref{fig:beta} summarizes the results for the continuum-extrapolated discrete $\beta$ function both with $c=0.25$ and 0.3. The once- and twice-smeared results are consistent and predict a $\beta$ function that is significantly smaller than the perturbative universal 2-loop value. Our results are even below the (non-universal)  4-loop \MSbar prediction, though the non-perturbative results increase monotonically in the range accessible in our simulations. 

It would be very interesting to repeat this calculation with  Wilson or domain wall fermions. Those actions might be able to probe stronger couplings and reach the expected chirally broken regime. Testing universality between different fermion formulations would be important as well.

\section{Conclusion}
In this paper I described several applications of the   ``t-shift" improved  gradient flow coupling. I illustrated this approach  for scale setting
 on 2+1+1 flavor HISQ configurations and  showed that after optimization  the ratios $\sqrt{t_0}/w_0$ or  $\sqrt{t_0}/r_1$ show negligible dependence on the lattice spacing.

The t-shift improved coupling reduces $\cO(a^2)$ lattice artifacts of the discrete $\beta$ function, improving the accuracy of continuum extrapolations in step scaling studies.  Our results in the $N_f=12$ SU(3) system  predict an  infrared conformal fixed point.  With 8 flavors we could follow the discrete $\beta$ function  up to  $\gtc(L) \approx 14$. We found that the  discrete $\beta$ function, while significantly smaller than the 2-loop prediction, increases  monotonically up to this gauge coupling.

\section*{Acknowledgements}
 I am indebted to Nathan Brown who graciously shared his gradient flow data with me. I thank my co-authors of \refcite{Cheng:2014jba,Hasenfratz:2014rna} for  fruitful collaborations.
 This research was partially supported by the U.S.~Department of Energy (DOE) through Grant Nos.~DE-SC0010005. Numerical calculations were carried out on the University of Colorado  Janus cluster partially funded by U.S.~National Science Foundation (NSF) 
 at Fermilab under the auspices of USQCD supported by the DOE; and at various NSF Computing Center through XSEDE allocations. 


{\small
\bibliography{gradflow_scale}
\bibliographystyle{apsrev4-1}
}



\end{document}